\documentclass[preprint,onecolumn,floats,floatfix,amsmath,nofootinbib,superscriptaddress]{revtex4-1}
\usepackage{graphicx,array,dcolumn}
\usepackage{calc,tabularx, epsfig}
\usepackage{hyperref}

\usepackage{amsmath}
\usepackage{amssymb}
\usepackage{multirow}
\usepackage{xspace}
\usepackage{slashed}
\allowdisplaybreaks[1]
\newlength{\figurewidth}
\newcommand{\beq}{\begin{equation}}
\newcommand{\eeq}{\end{equation}}
\newcommand{\bea}{\begin{eqnarray}}
\newcommand{\eea}{\end{eqnarray}}
\newcommand{\ba}{\begin{array}}
\newcommand{\ea}{\end{array}}

\newcommand{\mn}{{\mu\nu}}

\newcommand{\pt}{\partial}

\newcommand{\al}{\alpha}
\newcommand{\bt}{\beta}
\newcommand{\g}{\gamma}
\newcommand{\ep}{\epsilon}
\newcommand{\ta}{\theta}

\newcommand{\Lam}{\Lambda}

\newcommand{\D}{\Delta}

\newcommand{\om}{\omega}
\newcommand{\sg}{\sigma}

\makeatother
\begin{document}
%
\title{Signs and Stability in Higher-Derivative Gravity}
\setlength{\figurewidth}{\columnwidth}
%
\author{Gaurav Narain}
\email{gaunarain@itp.ac.cn}
\affiliation{
CAS Key Laboratory of Theoretical Physics,
Institute of Theoretical Physics (ITP), 
Chinese Academy of Sciences (CAS), Beijing 100190, P.R. China.}
%
%
\begin{abstract}
Perturbatively renormalizable higher-derivative gravity in four 
space-time dimensions with arbitrary signs of couplings has been considered. 
Systematic analysis of the action with arbitrary signs of couplings 
in lorentzian flat space-time for no-tachyons, fixes the signs.  
Feynman $+i\ep$ prescription for these sign further grants necessary 
convergence in path-integral, suppressing the field modes with large action. 
This also leads to a sensible wick rotation where quantum computation can be 
performed. Running couplings for these sign of parameters makes 
the massive tensor ghost innocuous leading to a stable and ghost-free 
renormalizable theory in four space-time dimensions. The theory has a 
transition point arising from renormalisation group (RG) equations, where the coefficient of
$R^2$ diverges without affecting the perturbative quantum field theory. 
Redefining this coefficient gives a better handle over the theory around the transition point.
The flow equations pushes the flow of parameters across the transition point. 
The flow beyond the transition point is analysed using the one-loop RG equations
which shows that the regime beyond the transition 
point has unphysical properties: there are tachyons, the path-integral loses positive 
definiteness, Newton's constant $G$ becomes negative and large, 
and perturbative parameters become large. These shortcomings 
indicate a lack of completeness beyond the transition point
and need of a non-perturbative treatment of the theory beyond the 
transition point.
\end{abstract}

\maketitle
%
%

\section{Introduction}
\label{intro}

Quantum field theory (QFT) is a beautiful framework established 
to address some of the mysteries of nature. Its success lies in the 
fact that it elegantly explains most of the observation 
seen in the accelerator experiments and condensed matter systems. 
On the other hand general relativity (a theory of gravity) has been 
widely used to understand and explain the mysteries at large scales. 
It comes to a puzzling situation when the two can't be easily
combined in a single theory without leading to problems. One of the 
most important problem that arises when methods of QFT are applied 
to Einstein-Hilbert gravity is that the resulting theory is plagued with 
ultraviolet (UV) divergences \cite{tHooft1974,Deser19741,Deser19742,
Deser19743,Goroff1985,Goroff1985a,vandeVen1991}, leading 
to non-renormalizablity. 

It has been noticed that the sickness of non-renormalizability can be 
cured by inclusion of higher-derivative terms \cite{Stelle19771,Stelle19772}, 
when the QFT of modified theory becomes renormalizable to all loops in four 
space-time dimensions. The path-integral of this theory is given by,
\begin{equation}
\label{eq:pathint}
Z
= \int \, {\cal D} \g_\mn
e^{i S_{\rm GR}}\, ,
\end{equation}
where
\beq
\label{eq:hdgact}
S_{\rm GR} 
= \int \frac{{\rm d}^4x \sqrt{-\g}}{16 \pi G} \biggl[
2 \Lam -a R + \frac{\omega R^2}{6 M^2}
- \frac{R_{\mu\nu}R^{\mu\nu} - 
\frac{1}{3}R^2}{M^2}
\biggr] \, .
\eeq
Here $R$ is the Ricci scalar and $R_{\mu\nu}$ is the 
Ricci tensor for the corresponding quantum metric $\g_\mn$. 
The last term is also proportional to square 
of Weyl tensor ($C_{\mn\rho\g}C^{\mn\rho\sg}$) in $3+1$ space-time 
dimensions (modulo total-derivative), $\omega$
is dimensionless and $M$ has dimension of mass,
$G$ is the gravitational Newton's constant and $\Lam$ is the cosmological 
constant term. The dimensionless parameter $a$ is for moment kept 
arbitrary but can take value either $+1$ or $-1$. 
It is a redundant parameter but is helpful in 
fixing some signs as will be seen later. 
In principle the path-integral in eq. (\ref{eq:pathint}) 
will also contain ghost action which enters due to gauge-fixing 
of the diffeomorphism invariant measure of theory, but for the purpose of
this paper will not be considered explicitly, though 
their contributions has been taken into account for the 
renormalisation group running of parameters 
\cite{Tomboulis:1980bs,Fradkin1981,Fradkin1982,Barth1983,Avramidi1985}.
The interesting thing to note here is that 
the UV renormalizability of the theory doesn't restrict the sign of 
various coefficients of terms in action \cite{Stelle19772}.

Higher-derivative theory has been studied many times in past.
This theory is notorious for its unitarity issues created 
by presence of ghost and tachyons \cite{Tomboulis:1977jk}. These were first 
studied in \cite{Salam1978,Julve1978,Tomboulis:1983sw,Antoniadis:1986tu} where proposal 
for avoiding them was made. After few years a euclidean 
version of the theory became popular as it was shown to 
have asymptotic freedom \cite{Fradkin1981, Fradkin1982,
Barth1983,Avramidi1985}. Over the years its coupling with matter has 
been investigated in euclidean framework \cite{Buchbinder1989,
Shapiro1989,Odintsov1989,Elizalde1994,Elizalde1995_1,
Elizalde1994_2,Buchbinder1992}. Later effect of Gauss-Bonnet 
was also studied \cite{deBerredoPeixoto2003,deBerredoPeixoto2004}.
But in general higher-derivative theories should not always be thought 
of having unitarity problems as has been shown in example 
considered in \cite{Smilga2004,Smilga2005,Donoghue2017}.

Recently, higher-derivative gravity has been studied in 
four space-time dimensions in lorentzian signature for issues of 
ghost and tachyons \cite{NarainA1, NarainA2}, while 
gauge-field coupling was investigated in detail in
\cite{NarainA3,NarainA4}. In these papers it was shown 
that unitarity problem can be tackled in the fully 
quantum theory by keeping the ghost mass always above the 
energy thus not allowing it to enter physical spectrum. 
The scale-invariant analog of this theory has  
been studied from phenomenological perspective in 
\cite{Strumia1,Salvio2016,Kannike2015}, where a scale 
dynamically arises in Einstein-frame of the theory. 
Fourth-order quantum-mechanical system were studied for unitarity issues 
in \cite{Salvio2015,Raidal2016}, where the authors argued that 
field theory when constructed in similar fashion for 
complicated gravitational systems might resolve unitarity 
problem. The idea of dimensional transmutation in scale-invariant 
theories where the scale arises via symmetry breaking 
due to quantum corrections ala Coleman-Weinberg 
has been studied in \cite{Einhorn2014,Jones2015,
Jones1,Jones2}. By making an analogy with QCD 
these systems were studied to get a resolution for the 
ghost problem using the wisdom acquired from 
non-perturbative sector of QCD \cite{Holdom2015,Holdom2016}.
Recently an interesting proposal has been made in 
\cite{Narain2016} where the scale-invariant higher-derivative theory 
studied directly in lorentzian was shown to break scale-symmetry 
via quantum corrections and in turn resolve problem of ghost
by choosing RG trajectories where the induced ghost mass is 
always above energy. 

These studies were conducted in perturbative framework 
(euclidean and lorentzian), however interesting 
developments have taken place in the field of 
asymptotic safety scenario \cite{Percacci2007} where the theory was 
considered using functional renormalisation group 
in euclidean signature \cite{Floreanini1993,Floreanini1994,
Codello2006,Codello2008,Groh2011,Ohta2013,Ohta2015,Ohta2016,Hamada2017}.
It was found that the theory admits spectral positivity 
\cite{Niedermaier2009}, where ghost can be tackled 
\cite{Floreanini1993,Floreanini1994} at the nontrivial 
fixed point \cite{Benedetti20091,Benedetti20092}.
However these analysis differ from the past ones in the sense that 
the theory is analysed at the non-gaussian fixed point in 
euclidean framework. 

The starting point for having a well-defined QFT is
the existence of a stable vacuum on which tower of states 
can be constructed and the states to have a positive norm 
so to have a unitary evolution. A QFT satisfying these 
basic requirements along with renormalizability is a 
well-defined QFT in which testable predictions can be made 
and meaningful computation can be performed.
The existence of a stable vacuum is guaranteed if the 
theory doesn't have any tachyons and stays in a regime where 
vacuum never become unstable. In case of higher-derivative 
gravity both these basic requirements gets challenged. 
If the parameters in the theory doesn't have appropriate signs
they it is noticed that the theory will have tachyons indicating that 
the vacuum of this theory is unstable. Therefore a QFT constructed 
with this is not reliable. However, there exits certain signs of 
parameters where there are no tachyons in four space-time 
dimensions in lorentzian signature 
(see for a similar study in three dimensions \cite{Ohta2011}).
In this short paper it is shown how 
one obtains these set of signs of parameters by carefully analysing the 
propagator for the tachyons. For these set of signs it is further 
noticed that the QFT constructed following feynman 
$+i\ep$ prescription has a necessary convergence 
suppressing field modes with large action. 

The renormalisation group flow of parameters is analysed in detail.
It is seen that the flow of the coefficient of $R^2$ terms goes 
to infinity, where the energy-scale reaches a maximum value
below which there are no tachyons and theory remains 
ghost free \cite{NarainA1, NarainA2, NarainA3, NarainA4}. 
This is a transition point. 
In this paper the theory around and beyond this transition 
point is explored to understand the true nature of this 
transition point and nature of the regime beyond.

The paper is organised as follows: section \ref{prop} deals 
with the analysis of propagator of theory in flat space-time 
where the tachyon analysis is done, section \ref{wickrot}
deals with a short introduction of the idea of wick rotation 
and $i\ep$ prescription where it is shown how the chosen sign 
of coupling allows convergence of lorentzian path-integral, 
section \ref{beta} deals with beta-functions and their analysis, 
finally conclusions with discussions 
are presented in section \ref{conc}.

\section{Propagator}
\label{prop}

Here we consider the propagator of theory given by the action 
in eq. (\ref{eq:hdgact}). For this we consider the fluctuations of 
metric around flat space-time $\g_\mn = \eta_\mn + h_\mn$
(with $\eta_\mn=\{+,-,-,-\}$). 
Flat space-time is not a solution of equation of motion for the 
action of theory given in eq. (\ref{eq:hdgact}), so in the 
following we will put $\Lam=0$ in the action. This is also 
justified as the renormalizability of the theory has been only 
demonstrated rigorously for $\Lam=0$ case only \cite{Stelle19772},
where it is possible to give a sensible particle content and 
use the methods/techniques of flat space-time quantum field theory.
Moreover this is further justified when one is interested in 
investigating high-energy behaviour, where 
the usage of flat space-time is rightfully advocated. 
Such flat space-time analysis have limitation 
which become relevant in deep infrared and are not 
a concern at short distances.

In the Landau gauge ($\pt_\mu h^\mn=0$), a physical gauge 
allowing only transverse modes to propagate, the propagator 
of metric fluctuation field $h_\mn$ is given by
\cite{NarainA1,NarainA2,NarainA3,NarainA4},
\beq
\label{eq:propHD}
\D^{\mn\al\bt} = (i16 \pi G) \biggl[
\frac{-2 M^2 P_2^{\mn\al\bt}}{q^4 - a M^2 q^2 }
+ \frac{M^2/\om P_s^{\mn\al\bt}}{q^4 - a M^2/\om q^2}
\biggr] \, ,
\eeq
where $q$ is the four-momentum of fluctuating field $h_{\mu\nu}$. 
Various spin projectors are  
$P_2^{\mn\al\bt} = 
\frac{1}{2} \left[ T_{\mu\alpha} T_{\nu\beta} + 
T_{\mu\beta}T_{\nu\alpha} \right] 
- \frac{1}{3} T_{\mu\nu}T_{\alpha\beta} $, 
$P_s^{\mn\al\bt} = 
\frac{1}{3} T_{\mn} \, T_{\al\bt}$, 
where $T_{\mn}=\eta_\mn - q_{\mu}q_{\nu}/q^2$.
They project the spin-2 and spin-0 component 
of the $h_\mn$ field. Each of the denominator is a quadratic 
polynomial in $q^2$ ($=\eta_\mn q^\mu q^\nu$), 
implying two roots (real or complex 
depending on sign of parameters). 
One can do partial fraction after factorisation of 
denominators to write the propagator in simple form. 
\beq
\label{eq:lam0prop}
\D^{\mn\al\bt} = \frac{(i16 \pi G)}{a} \biggl[
\frac{(2P_2^{\mn\al\bt} -P_s^{\mn\al\bt})}{q^2}
- \frac{2P_2^{\mn\al\bt}}{q^2 - a M^2}
+ \frac{P_s^{\mn\al\bt}}{q^2 - aM^2/\om} 
\biggr] \,.
\eeq
So far the propagator is just a function of $q^2$ where the 
metric dependence enters implicitly. 
This propagator consist of just three parts: the usual 
massless graviton, massive tensor mode, and massive 
scalar mode. The massless graviton and massive scalar 
has the same sign of propagator while the massive tensor 
has the opposite sign. By comparing with low-energy, 
known Einstein-Hilbert gravity, one can fix the sign of 
parameter $a=+1$ (where $G$ is taken to be 
positive for attractive gravity). 
This is mandatory otherwise the 
gravitational-wave in flat space-time will have negative 
norm, but this can be actually absorbed in the definition 
of $G$ which decides how gravity will couple with matter. 
The parameter that will decide how at low energy matter 
couples with gravity is $a/G$, which should be positive 
for attractive nature of gravity. 
For consistency we will keep $a=+1$. Once this is 
fixed we have freedom to fix the sign of $M^2$ and $\om$.
This is done by requiring that the propagator shouldn't have 
any tachyons. For the signature given by $\eta_\mn$ the
sign of $M^2$ and $\om$ should be positive to have no tachyons. 
Reversing the signature sign will result in change in sign of 
terms where to avoid tachyons $M^2 \to -M^2$ while $\om$ remains positive. 
This information can be neatly written in a tabular form.
This is given in table. \ref{tab:inval}.

\begin{table*}[t]
\begin{tabular}{| c | c | c | c | c | c |}
\hline 
Signature & Spin & $M^2>0$, $\om>0$ & $M^2<0$, $\om>0$
& $M^2>0$, $\om<0$ & $M^2<0$, $\om<0$ \\
\hline \hline
\multirow{ 2}{*}{$\eta_\mn$} & Spin-2 & 
No & Yes & No & Yes \\
 & Spin-s & No & Yes & Yes & No \\
 \hline
\end{tabular}
\caption{
Tachyon analysis for the signature and sign of couplings.
Here yes and no refers to tachyon presence and absence 
respectively. 
}
\label{tab:inval}
\end{table*}

An alternative possibility is considering $a=-1$ and reversing 
the sign of $G$, so that $a/G$ remains unchanged. This will 
give rise to tachyons in massive tensor and massive scalar
mode, thereby demanding $M^2\to-M^2$ for tachyon elimination. 
However the action of two cases $a=\pm1$ is same. 
In this sense parameter $a$ is redundant. 
In either case these are different from the version 
of lorentzian theories considered in 
\cite{Holdom2015,Holdom2016}, where the 
propagator has tachyons. 

\section{Wick rotation}
\label{wickrot}

Here in this section we discuss about standard wick-rotation and 
feynman $i\ep$ prescription. The first part of which is a small 
review of textbook material covered here to make the paper more clear. 
The sign of the $i\ep$-prescription
gets fixed from the definition of the starting point path-integral. 
For the integrand in path-integral
$e^{\pm iS}$ the prescription is $\pm i\ep$ to make the 
path-integral convergent. This allows for suppression of those 
field modes for which the action is large. Systematically this is 
achieved by doing the following $\int {\rm d}t L \to \int {\rm d}t(1 + 
i \ep) L = S + i\ep S$ ($0<\ep\ll1$), where the Lagrangian $L=\int {\rm d}^3x \mathcal{L}$
and $\mathcal{L}$ is the Lagrangian density. Here 
the convergence is achieved by doing transformation 
$t \to t+ i\ep$, which is a standard practice in defining lorentzian 
path-integrals which also respects all the symmetry of the original action.
Moreover, this also leads to propagator with shifted poles 
in a natural manner, a standard thing covered in 
most textbooks on quantum field theory. 

In the path-integral the $+i\ep$ offers a 
suppression factor for field modes for which the action is large. 
In our particular case this will imply
\beq
\label{eq:pathintS}
Z
= \int \, {\cal D} \g_\mn
e^{i S_{\rm GR} - \ep S_{\rm GR}}\, ,
\eeq
where $S_{\rm GR}$ is given by eq. (\ref{eq:hdgact})
and $\g_\mn$ is the quantum metric. 
To have the required convergence and well-defined path-integral 
it is required that the coefficient of the $R^2$ term to be positive
{\it i.e.} $\om>0$. In the $+i\ep$ prescription this will give 
rise to an additional term in exponent: $-\ep \om R^2$.
For field modes with large $R^2$, such a term will heavily 
suppress that mode resulting in convergence of the 
lorentzian path-integral (in euclidean path-integral a positive 
coefficient of $R^2$ is needed for positive definiteness 
\cite{Gibbons1978,Macrae1981,Fradkin1982}). Interestingly $\om>0$ is 
also the regime which avoids tachyons in lorentzian signature 
(both for $\pm\eta_\mn$). 

In momentum space the prescription shifts the locations of poles 
in the flat space-time propagator\footnote{In momentum space the quadratic part of the any 
action will look like $(1+i\ep)(q^2 - m^2) = q^2 - m^2 + i\ep^\prime$,
where $\ep^\prime = \ep (q^2-m^2)$ is the new parameter. 
This will result in standard $+i\ep$ prescription.}
In the higher-derivative case for $a=+1$, $\Lam=0$
and signature $\eta_\mn$ the propagator will be,
\beq
\label{eq:lam0propiep}
\D^{\mn\al\bt} = (i16 \pi G) \biggl[
\frac{(2P_2^{\mn\al\bt} -P_s^{\mn\al\bt})}{q^2 + i\ep}
- \frac{2P_2^{\mn\al\bt}}{q^2 - M^2 + i\ep}
+ \frac{P_s^{\mn\al\bt}}{q^2 - M^2/\om + i\ep} 
\biggr] \,.
\eeq
In a complex $q_0$ plane the $+i\ep$ prescription will 
shift the poles on the real axis to second and fourth quadrant. 
When performing loop-computations for the quantum corrections,
this will allow to choose contour which doesn't enclose these shifted 
poles. The integral along the contour in the end will reduce to 
replacement of $q_0 \to iq_0$. This is the standard 
wick rotation in flat space-time usually covered 
in textbooks. It shows how the wick rotation is tied to 
$i\ep$ prescription whose sign is chosen to provide 
appropriate convergence in the path-integral. 

In the case when the poles lie on imaginary axis, 
which happens when there are tachyons, the $+i\ep$
prescription shifts the poles in second and fourth 
quadrant. Here the same contour is fine and 
usual wick rotation can be done as the shifted poles are 
outside contour. In the present case however where appropriate 
choice of signs evades tachyons such a thing is not needed.

\section{Beta-functions and RG analysis}
\label{beta}

The perturbative field theory is constructed in 
parameters $M^2G$ and $M^2G/\om$, where loop 
computation are performed and beta-functions are computed. 
These beta-functions have been computed earlier 
\cite{Avramidi1985,Avramidi2000} in the case of 
euclidean field theory. In \cite{NarainA1, NarainA2} the 
beta-functions were translated to lorentzian signature 
in Landau gauge. Here we take over those beta-functions
for further analysis. These are given by,
\bea
\label{eq:betaM2G}
&& \frac{{\rm d}}{{\rm d} t} \left( \frac{1}{M^2 G} \right)
= - \frac{133}{10 \pi} \, ,
\\
\label{eq:beta_wM2G}
&& \frac{{\rm d}}{{\rm d} t} \left(
\frac{\omega}{M^2 G} \right) =
\frac{5}{3 \pi} \left( \omega^2
+ 3 \omega + \frac{1}{2} \right) \, ,
\\
\label{eq:betaG}
&& \frac{{\rm d}}{{\rm d} t} \left(\frac{1}{G} \right)
= \frac{5M^2}{3 \pi} \left(\omega
- \frac{7}{40 \omega} \right) \, .
\eea
where $t= \ln (\mu/\mu_0)$ and the {\it r.h.s.} for all beta-function contain
the leading contribution in $G$ ($M^2G$ is also taken to be small), 
with higher powers coming from higher loops being neglected. 
It should be mentioned that the RG flow of coupling $G$ is gauge-fixing 
dependent. Here its flow is given in Landau gauge \cite{Avramidi2000,NarainA1,NarainA2}. 
Also, we have ignored the cosmological constant term and its running, which is 
done so that flat space-time exists as a solution to equation of motions 
on whose background quantum field theory analysis can be trustfully 
performed. From this we first extract the flow of parameter $\om$, which is the 
coefficient of the $R^2$ term and for no-tachyons needs to be 
positive. Its running is given by,
\beq
\label{eq:beta_w}
\frac{{\rm d} \om}{{\rm d} t} 
= \frac{5 M^2 G}{3 \pi} 
\left(\om^2 + \frac{549}{50} \om + \frac{1}{2} \right) 
= \frac{5 M^2 G}{3 \pi}  (\om + \om_1)(\om +\om_2)
\eeq
where $\om_{1,2} = (549 \mp \sqrt{6049})/100$. 
Both the roots $-\om_1$ and $-\om_2$ lie in the tachyonic 
regime. The fixed point given by $-\om_1$ is repulsive 
while $-\om_2$ is attractive. Since the {\it r.h.s.} of beta-function 
of $\om$ is positive and increase with $\om$, therefore 
$\om$ is a monotonic increasing function of RG time $t$. 
No-tachyon condition puts a constraint on the allowed range of 
$\om$ to be $0\leq \om \leq \infty$. One can solve the 
flow of $\om$ in terms of RG time $t$ and obtains 
\beq
\label{eq:tw}
t= T
\Biggl[
1- 
\left( \frac{\omega + \omega_2}{\omega + \omega_1}
\cdot 
\frac{\omega_0 + \omega_1}{\omega_0 + \omega_2} \right)^{\alpha}
\Biggr] \, ,
\eeq
where $T=10 \pi/(133 M_0^2 G_0)$ 
and $\alpha= 399/(50(\omega_2 -\omega_1))>0$, with
subscript $0$ meaning that the coupling parameters are 
evaluated at $t=0$ or $\mu = \mu_0$, which will be decided later. 
The bound on $\om$ translates to a lower and upper bound on 
value of RG time $t$ within which the parameter $\om$ 
remains positive and hence avoid tachyons. 
The lower bound $t_{\rm min}$ and upper bound 
$t_{\rm max}$ are given by,
\beq
\label{eq:tmax_min}
\frac{t_{\rm min}}{T} = 
1 - \left(\frac{\omega_2}{\omega_1} 
\frac{\omega_0 + \omega_1}
{\omega_0 + \omega_2}
\right)^{\alpha} \, ,
\hspace{5mm}
\frac{t_{\rm max}}{T} = 
1 - \left(
\frac{\omega_0 + \omega_1}
{\omega_0 + \omega_2}
\right)^{\alpha} \, ,
\eeq
respectively. Here $t_{\rm min}$ is in infrared while 
$t_{\rm max}$ is in UV. As $t\to t_{\rm max}$, $\om\to\infty$. 
The emergence of $t_{\rm max}$ is puzzling as the theory has 
been studied without imposing any cutoff. This bound 
arises from the RG flow equations due the restriction 
imposed by stability of vacua (no-tachyons) in lorentzian signature. 
Although at this point the parameter $\om\to\infty$ the perturbation 
theory remains valid as the parameters $M^2G$ and $M^2G/\om$ 
remain small, with the amplitudes computed using them 
remains well-defined. It is therefore natural to go beyond this 
bound and curiously explore the regime beyond $t_{\rm max}$. 
In this paper we do this exploration by extrapolating the 
beta-functions in this regime and the findings are 
presented in this paper. 

A good way to explore this large $\om$ regime is to define a new 
parameter $\ta=1/\om$. Unlike $\om$, this new parameter $\ta$
is continuous near $\ta=0$ ($\om=\infty$). This new parametrisation 
allows to study the flow equation near this singular-point 
in a systematic manner. In particular the behaviour of flows 
on both sides of $\ta=0$ point is different. The regime dictated 
by $t<t_{\rm max}$ is $\ta>0$, while the regime for $t>t_{\rm max}$
is given by $\ta<0$. In the following subsections 
both these regimes will be studied, and will be realised that 
$\ta=0$ is a transition point where a smooth crossover happens 
from one regime to another. 

\subsection{$\ta>0$ Regime}
\label{ta0P}

In this regime the beta-function of $\ta$ is given by following, 
\beq
\label{eq:thetabeta}
\frac{{\rm d} \ta}{{\rm d} t} 
= - \frac{5 M^2 G}{3 \pi} 
\left(1 + \frac{549}{50} \ta + \frac{\ta^2}{2} \right) \, .
\eeq
This also has two fixed points $-\ta_1 = -1/\om_1$ and 
$-\ta_2 = -1/\om_2$. Both of these lie in tachyonic 
regime (negative $\om$). But now $-\ta_1$ is repulsive while $-\ta_2$ is attractive 
fixed point. The parameter $\ta$ has a continuous 
behaviour around the point $\ta=0$, where its smooth 
flow is dictated by the first order ODE given in eq. (\ref{eq:thetabeta}).
At $\ta=0$ the beta-function of $\ta$ is negative thereby 
forcing $\ta$ to decrease further. It is expected as $\ta=0$ is 
not a fixed point for the flow of $\ta$. 

Other couplings parameters $M^2G$ and $G$ can then be solved 
in terms of continuos parameter $\ta$ to study their behaviour near and 
across the $\ta=0$ point.  
Writing the ODE for $M^2G$ in terms of $\ta$ quickly shows the 
solution of $M^2G$ in terms of $\ta$ to be,
\beq
\label{eq:M2GtaP}
M^2G = M^2_0G_0 \left(\frac{\ta+\ta_1}{\ta_0+\ta_1} 
\frac{\ta_0+\ta_2}{\ta+\ta_2}\right)^{-\al} \, ,
\eeq
where $\al$ is the same number 
appearing in eq. (\ref{eq:tw}) and $M^2_0G_0$ is the 
initial value of the parameter $M^2G$ at the reference 
point $\ta_0$ which will be fixed next. As $\ta \to0$, 
$M^2G$ increase and goes to a fixed value 
$M^2_0G_0(\ta_1(\ta_0+\ta_2)/(\ta_2(\ta_0+\ta_1)))^{-\al}$. 

The flow of $G$ can be similarly solved. The beta-function of 
$G$ can be expressed in terms of $\ta$ as,
\beq
\label{eq:betaGB}
\frac{{\rm d} G}{{\rm d} \ta} 
= \frac{2G\left(1 - \frac{7}{40} \ta^2\right)}{\ta(\ta+\ta_1)(\ta+\ta_2)} \, .
\eeq
The flow of parameter $G$ is interesting as it has 
various fixed points. 
From its beta-function we notice that it has three 
fixed points: two at $\ta=\pm\sqrt{40/7}$ and one where 
$G/\ta$ vanishes. 
The fixed point $\ta=-\sqrt{40/7}$ lies in the unphysical 
domain so can be ignored. The fixed point given by 
$\ta=\sqrt{40/7}$ is attractive fixed point, we call this 
a reference point $\ta_0$ (the point $\ta=-\sqrt{40/7}$ 
is also attractive in nature but lies in unphysical regime). 
The third fixed point where simultaneously both $G$ and 
$\ta$ vanishes is a little tricky to analyse. For this analysis 
we first solve analytically $G$ in terms of $\ta$. This is 
easily achieved as eq. (\ref{eq:betaGB}) is a first order ODE
with {\it r.h.s.} being ratio of polynomials. 
This is given by,
\beq
\label{eq:GtaP}
\frac{G}{G_0} = \left(\frac{\ta}{\ta_0}\right)
\left(\frac{\ta+\ta_1}{\ta_0+\ta_1}\right)^{A_1}
\left(\frac{\ta+\ta_2}{\ta_0+\ta_2}\right)^{A_2} \, ,
\eeq
where 
\beq
\label{eq:A1A2}
A_1 = \frac{2(1-\frac{7}{40} \ta_1^2)}{\ta_1(\ta_1-\ta_2)}\, , 
\hspace{5mm}
A_2 = -\frac{2(1-\frac{7}{40} \ta_2^2)}{\ta_2(\ta_1-\ta_2)} \, .
\eeq 
From this we see that as $\ta\to0$, $G/G_0\sim \ta$ and 
approaches zero therefore this point $\ta=0$ is 
a fixed point for the flow of $G$, where $G$ also 
vanishes. Taking second derivative of eq. (\ref{eq:betaGB})
it is seen that in the limit $\ta\to0$ it has negative value 
thereby implying that it is an attractive fixed point for 
flow of $G$. However it is not a fixed point for the 
parameter $\ta=1/\om$. The RG flow of $\ta$ will therefore 
push $\ta$ beyond this point and take it in the negative regime. 
The point $\ta=0$ is then just a crossover point where the 
transition from $\ta>0$ to $\ta<0$ happens. Interestingly 
during this transitory phase perturbation theory remains valid
but $G$ changes sign becoming a repulsive interaction. 
Although $G=0$ (at $\ta=0$) is an attractive fixed point for $G$, 
but it is dragged away from this point as it is not a fixed point for $\ta$. 
As $\ta$ is pushed into negative regime it forces $G$ 
to change sign. Physically this will imply that the gravitational interaction 
between two bodies beyond a certain length scale becomes 
repulsive. In next subsection we study the theory after cross over.

\subsection{$\ta<0$ regime}
\label{taM}

The parameter space beyond the crossover point $\ta=0$
is dictated by $\ta$ becoming negative where the 
Newton's constant $G$ goes to negative values. This 
regime lies at ultra-high energies. 
To explore this regime systematically we write $\ta=-\bt$. 
The beta-function of the parameter $\bt$ is given by,
\beq
\label{eq:betaB}
\frac{{\rm d} \bt}{{\rm d} t} 
=  \frac{5 M^2 G}{3 \pi} 
\left(1 - \frac{549}{50} \bt + \frac{\bt^2}{2} \right) \, .
\eeq
This regime is governed by the parameter $\bt$. 
The flow equation of $\bt$ shows that it has two 
fixed point $\bt_1=\ta_1$ and $\bt_2=\ta_2$
where $\bt_1<\bt_2$. 
It is seen that $\bt_1$ is attractive and $\bt_2$ 
is repulsive. Moving over the cross-over point 
it is expected that due to continuity the flow of $\bt$
will move towards the attractive fixed point located at $\bt_1$. 

The flow of other parameters can be studied in terms of 
$\bt$, where it is expected that running of $\bt$ to 
the attractive fixed point $\bt_1$ will have consequences 
for the flows of other coupling parameters. To investigate 
this throughly first the beta-function of $G$ is re-expressed 
in terms of $\bt$. This is given by,
\beq
\label{eq:betaGB1}
\frac{{\rm d} G}{{\rm d} \bt} 
= \frac{2G\left(1 - \frac{7}{40} \bt^2\right)}{\bt(\bt-\bt_1)(\bt-\bt_2)} \, .
\eeq 
This first order ODE can be solved easily as before 
giving the solution to be,
\beq
\label{eq:GtaP1}
\frac{G}{G_0} = -\left(\frac{\bt}{\ta_0}\right)
\left(\frac{\bt_1-\bt}{\ta_0+\bt_1}\right)^{A_1}
\left(\frac{\bt_2-\bt}{\ta_0+\bt_2}\right)^{A_2} \, ,
\eeq
where $-\ta_0$ is the previous reference point and $G_0$ is 
the value of the corresponding point. This gives the evolution of $G$ 
in the regime $\ta<0$. The flow shows that at the 
crossover point $\ta=0$ the Newton's constant $G$ is zero 
and make the transition smoothly. 
Beyond the crossover point Newton's constant $G$ 
becomes negative. This means that in this domain 
gravitational interaction will be repulsive. 
It is further seen that after the crossover as the flow 
of $\bt$ approaches the attractive fixed point at $\bt_1$,
the flow of $G$ diverges ($G\to-\infty$), implying that 
at this point the gravitational repulsive force will be infinite.
 
The flow of parameter $M^2G$ can be solved in terms of 
$\bt$ using the running of $M^2G$ and $\bt$
given in eq. (\ref{eq:betaM2G}) and (\ref{eq:betaB})
respectively. This is given by,
\beq
\label{eq:M2Gbt}
M^2G = M^2_0G_0 \left(\frac{\bt-\bt_1}{\ta_0+\bt_1} 
\frac{\ta_0+\bt_2}{\bt-\bt_2}\right)^{-\al} \, .
\eeq
From this flow we see that at the crossover point 
$M^2G$ makes a smooth transition and remains 
small. However as $\bt$ approaches the attractive 
fixed point $\bt_1$, the parameter $M^2G$
diverges as $\al>0$.

\section{Conclusion}
\label{conc}

In this short paper we investigated UV renormalizable fourth-order 
higher-derivative gravity in four space-time dimensions in lorentzian signature
with arbitrary parameters. We study the nature of flat space-time 
propagator for various signs of parameters and find the set of signs 
for which there exist no-tachyons. Interestingly in this domain of 
parameters with these signs it is noticed that in the 
feynman $+i\ep$ prescription, the path-integral 
has the required convergence suppressing field modes with large action. 
This is the same regime where there are no tachyons. The RG 
flows of the parameters are analysed extensively. The parameter 
$\om$ (coefficient of $R^2$) has a range between zero and infinity 
within which the theory remains tachyon free. This bound translates 
into a bound on the allowed energy domain where theory remains 
tachyon-free. In this paper we analyse what happens beyond the 
upper bound of the RG time $t_{\rm max}$, where 
the parameter $\om$ runs to infinity. The beta-functions are analysed for 
large $\om$ regime by defining the parameter $\ta=1/\om$. 
The parameter $\ta$ has a well-defined behaviour 
in the large $\om$ regime. The regime below $t_{\rm max}$
is then given by $\ta>0$, while the regime beyond is 
dictated by $\ta<0$. The two regimes are analysed 
separately. 

In the regime $\ta>0$ it is seen that the flow of $\ta$ 
approaches zero logarithmically (in terms of energy). 
Moreover, the flow of $G$ also approaches zero as $\ta\to0$. 
But the ratio $G/\ta$ remains finite at this point. This point 
is an attractive fixed point for $G$, however it is not a
fixed point for parameter $\ta$. As a result the flow is 
dragged away from this point. The non-zero beta-function of 
$\ta$ at the point $\ta=0$ pushes $\ta$ to negative values, 
making the point $\ta=0$ a transition point where crossover 
happens. At this crossover perturbation theory is well-defined.
This crossover also changes the sign of $G$,
thereby making it negative in the regime $\ta<0$. 

The regime beyond $t_{\rm max}$, given by $\ta<0$ has 
different flow of couplings. This regime is studied 
in terms of $\bt=-\ta$. In this regime it is seen 
that once beyond the crossover point, the flow of $\bt$
quickly approaches the attractive fixed point $\bt_1$. The 
Newtons' constant $G$ beyond the crossover point 
becomes negative and is dragged to the attractive fixed point 
of $\bt$ where it diverges. Negative $G$ implies that the 
gravitational force becomes repulsive in this regime 
and this repulsion is infinite at the fixed point. 

The theory considered and analysed here in this paper 
(and previous works \cite{NarainA1, NarainA2,Narain2016})
is different from the studies conducted by other authors  
\cite{Fradkin1981, Fradkin1982,
Barth1983,Avramidi1985,Buchbinder1989,Shapiro1989,Odintsov1989,Elizalde1994,
Elizalde1995_1,Elizalde1994_2,Buchbinder1992,deBerredoPeixoto2003,
deBerredoPeixoto2004}, which were done in 
euclidean signature where the signs of couplings 
were taken to get positive definiteness for euclidean 
path-integral. Those choice of signs further gives them 
asymptotic freedom for flow of couplings but theory contains 
tachyons. This was mentioned in their 
work \cite{Fradkin1981,Fradkin1982Avramidi1985}. 
However, positive-definite asymptotically free euclidean theories 
considered by them are is correct in those signatures
(modulo the tachyon problem which is important in 
infrared) but results obtained from them should not be 
expected to hold true for lorentzian theories.  

The current work (and previous works \cite{NarainA1, NarainA2,Narain2016}) 
differ from them in the sense we constructed the theory directly in 
lorentzian signature and signs are chosen so that tachyons are avoided.
This choice of signs also differ from the other work done in 
scale-invariant lorentzian theories where the signs 
were chosen to have asymptotic freedom
\cite{Strumia1}. The choice of signs in this paper 
(and past ones \cite{NarainA1, NarainA2,Narain2016})
further allows convergence in path-integral following feynman $i\ep$ prescription. 
Once the basic requirement of signs and stability is 
satisfied, the one-loop RG flows are investigated and 
conclusions are obtained.

In this lorentzian theory however the regime beyond $t_{\rm max}$ 
is unphysical and the theory goes in a non-perturbative domain. 
This is inevitable as at the turning point the beta-function of 
$\ta$ is non-zero pushing it to negative values. 
In this regime the parameter $\ta$ becomes negative indicating 
that tachyons are present implying the instability of vacuum. 
The change of sign of gravitational coupling $G$ is an interesting 
outcome leading to repulsive gravity, but it arises in a regime where there are 
tachyons and attachment of any useful meaning to it 
should be done with skepticism. Moreover in this regime 
the perturbative parameters of QFT, $M^2G$ and $M^2G/\om$
grows large. They becomes infinite at the attractive 
fixed point of $\bt$, thereby implying that this regime 
require a non-perturbative treatment. It will be interesting 
to see how these issues changes when the same analysis 
is performed with non-perturbative flow equations obtained 
through functional renormalisation group 
within the asymptotic safety scenario picture.

The shortcomings beyond $t_{\rm max}$, witnessed in the one-loop analysis
might indicate the following: (a) The one-loop analysis is not
reliable beyond $t_{\rm max}$ and a higher-loop study is required where $t_{\rm max}$ 
might be pushed to infinity. (b) The theory needs some extra
ingredient (like Strings etc) to make it UV complete, in which 
sense the theory is an effective theory. In fact the low energy 
limit of string theory does contain an infinite number of curvature-dependent 
terms (local and non-local) that goes beyond quadratic gravity, which
hints that perhaps the addition of further higher-derivative terms 
might bring better stability and convergence in path-integral. 
(c) Beyond $t_{\rm max}$ the effects of non-locality should 
be taken into account along the lines of super-renormalizable theories
\cite{Modesto2011,Biswas2011,Modesto2014,Tomboulis2015}.
(d) This might also be hinting that beyond the transition point 
the theory become entirely conformal \cite{tHooft2015,tHooft2016}.
The terms $R$ and $R^2$ are 
induced in low energy via spontaneous symmetry 
breaking of weyl-invariance due to radiative corrections
\cite{Adler19803,Zee1980,Adler1982,Nepomechie1983,Zee1983,Smilga1982}.
These possibilities are worthy of exploration in future. 

Although the shortcomings in this regime are worrisome 
but nevertheless within the large allowed energy range 
($t_{\rm min}\leq t \leq t_{\rm max}$) the theory remains 
renormalizable to all loops, tachyon and ghost-free \cite{NarainA1, NarainA2}.
It describes a well-defined renormalizable and unitary QFT
in this energy range. This regime is large enough to address 
length scales ranging from Planck to current
cosmological scales. Moreover, the existence of $t_{\rm max}$ 
is a one-loop effect which might possible go away (or $t_{\rm max}\to\infty$)
in higher-loop studies as mentioned in \cite{NarainA1, NarainA2}. Even 
in one-loop, by appropriate choice of parameters one can 
push $t_{\rm max}$ all the way to Planck-scale (and beyond). This is a relief as the 
renormalizable theory where ghosts and tachyons are eradicated can be 
trusted all the way to ultra high energies.

\bigskip
\centerline{\bf Acknowledgements} 

I would like to thank Nirmalya Kajuri and Tuhin Mukherjee 
for useful discussions at various stages of this 
work. I am grateful to KITPC and Prof. Tianjun Li for support.
I would like to thank Ghanashyam Date and IMSc, Chennai for support where 
a part of work was done.



\end{document}